\documentclass[a4paper,fleqn,usenatbib]{mnras}

\usepackage{amsmath}	% Advanced maths commands
\usepackage{amssymb}	% Extra maths symbols

\usepackage{mathptmx}
\usepackage{txfonts}

\usepackage[T1]{fontenc}
\usepackage{ae,aecompl}
\usepackage{float}

\usepackage{times}

\usepackage{graphicx}
\usepackage{epstopdf}
\usepackage{color}
\usepackage{bm}% bold math
\title[Quasar Disc Size with Wind]{Reconciling the Quasar Microlensing Disc Size Problem with a Wind Model of Active Galactic Nucleus}

\author[Li, Yuan, \& Dai]{
Ya-Ping Li$^{1}$\thanks{E-mail: leeyp2009@gmail.com (YPL)},
Feng Yuan$^{1,2}$\thanks{E-mail: fyuan@shao.ac.cn (FY)},
\& Xinyu Dai$^{3}$\thanks{E-mail: xdai@ou.edu (XD)}\\
$^{1}$Key Laboratory for Research in Galaxies and Cosmology, Shanghai Astronomical Observatory, Chinese Academy of Sciences, 80\\
Nandan Road, Shanghai 200030, China\\
$^{2}$School of Astronomy and Space Sciences, University of Chinese Academy of Sciences, No. 19A Yuquan Road, Beijing 100049, China\\
$^{3}$Department of Physics and Astronomy, University of Oklahoma, 440 W. Brooks Street, Norman, OK 73019, USA
 \\
}

\date{Accepted xxx. Received xxx; in original form xxx}

\pubyear{2018}

\begin{document}
\label{firstpage}
\pagerange{\pageref{firstpage}--\pageref{lastpage}}
\maketitle

\begin{abstract}
{Many analyses have concluded that the accretion disc sizes measured from the microlensing variability of quasars are larger than the expectations from the standard thin disc theory by a factor of $\sim4$. We propose a simply model by invoking a strong wind from the disc to flatten its radial temperature profile, which can then reconcile the size discrepancy problem.  This wind model has been successfully applied to several microlensed quasars with a wind strength $s\lesssim1.3$ by only considering the inward decreasing of the mass accretion rate (where $s$ is defined through $\dot{M}(R)\propto({R}/{R_{0}})^{s}$ ). After further incorporating the angular momentum transferred by the wind,  our model can resolve the disc size problem with an even lower wind parameter.
The corrected disc sizes under the wind model are correlated with black hole masses with a slope in agreement with our modified thin disc model.
}
\end{abstract}

\section{Introduction}
It is widely accepted that active galactic nuclei (AGNs) in the distant universe are powered by accretion discs around the supermassive black holes. The simple \citet{Shakura1973} thin-disc model (see also \citealt{Novikov1973}) remains the standard model for luminous AGNs due to its success in modeling some major features in observations (e.g., the ``Big Blue Bump" in the spectral energy distribution of quasars). However, this simple model has shown some difficulties in some aspects, e.g., in explaining the soft X-ray continuum, optical polarization, and variability \citep{Koratkar1999}.

More remarkably, a ``size problem" has been recently identified by many works. Based on the optical-UV microlensing observations for quasars, the disc size can be measured from microlensing variabilities. We can also obtain two disc sizes based on the thin disc theory (see Equation~\ref{eq:rth2}) and the magnification-corrected flux (see Equation~\ref{eq:rflux}). These observations show that the disc sizes measured from microlensing variabilities are systematically larger than the thin-disc theory size by a factor of $\sim4$ \citep[e.g.,][]{Pooley2007,Dai2010,Morgan2010,Jimenez2012,Blackburne2014,Munoz2016,Motta2017}.
Several suggestions have been put forward to reduce the size discrepancy, e.g., scattering a significant fraction of the disc emission or including the line emission contamination from larger physical scales \citep{Morgan2010}, a flatter disc temperature profile because of some unknown reasons \citep{Dai2010,Morgan2010,Bonning2013}, an inhomogeneous disc with large temperature fluctuations \citep{Dexter2011,Cai2018}, and a super-Eddington accretion disc with an optically thick envelope \citep{Abolmasov2012}.

In this work, we propose that a thin disc with wind can flatten the temperature profile, which then can solve the disc size problem. This shares a similarity with the suggestion of a flatter disc temperature profile mentioned above, although the physical origin of the flattening has not been linked to disc winds in previous works yet. Observationally, a flattening of temperature profile than the theoretical expectation of $3/4$ from the thin-disc model has been confirmed for several microlensing studies of disc structures \citep[e.g.,][]{Bate2008,Poindexter2008,Bate2018}. However, a few sources show that the temperature profile could be even steeper than $3/4$, although the uncertainty is still very large \citep[e.g.,][]{Eigenbrod2008,Jimenez2014,Munoz2016}.
Another possible problem for the standard thin disc model is the difficulty in reproducing the turnover at $\lambda\sim1000~{\AA}$ for the AGN spectral energy distribution (SED). This has been extensively studied by many works with a disc wind model \citep{Kuncic2007,Slone2012,Laor2014,Sun2018b} and uncorrected host galaxy extinction \citep{Capellupo2015}.

%and black hole growth rate of AGNs
There is now compelling evidence for the existence of wind in different types of accretion flows both observationally and theoretically. For the hot accretion flow, both hydrodynamic (HD) and magnetohydrodynamic (MHD) numerical simulations have found that the mass inflow rate decreases with decreasing radius (see review by \citealt{Yuan2014}). \citet{Yuan2012a} show that the inward decrease of the accretion rate is attributed to the significant mass loss through wind (see also \citealt{Narayan2012,Gu2015,Yuan2015}). This result is confirmed later by \emph{Chandra} observations for the supermassive black hole in our Galaxy \citep{Wang2013}. For the standard thin disc powering luminous quasars, numerous pieces of observational evidence have been accumulated via studies of broad absorption line quasars (e.g., \citealt{Arav2001,Chartas2003,Crenshaw2003,Dai2008,Tombesi2010,Dai2012,Tombesi2014,Gofford2015,King2015,Liu2015}). and emission line quasars \citep[e.g.,][]{Sun2018a}.
The launching location of the wind is within $100~R_{\rm g}$ \citep{Tombesi2012,Gofford2015,Tombesi2015} with the associated column densities of the absorbers being in the range of $10^{22}-10^{24}~{\rm cm^{-2}}$ \citep{Tombesi2011}.
It is generally believed that these winds are launched from the thin disc around the central black hole by radiation (e.g., \citealt{Shields1977,Proga2000,Risaliti2010,Nomura2016,Nomura2017}),
thermal
(e.g., \citealt{Begelman1983,Woods1996,Krolik1984,Chelouche2005,Everett2007}), magnetic mechanisms
(e.g., \citealt{Blandford1982,Contopoulos1994,Konigl1994,Fukumura2018,Kraemer2018}),
and a combination of them (e.g., \citealt{Proga2003,Waters2018}).

The paper is organized as follows. Our wind model is described in details in Section~\ref{sec:method}, and the influence of the angular momentum transfer by wind is further discussed in Section~\ref{sec:ang}. We apply our wind model to several microlensed quasars in Section~\ref{sec:results}.  The final section is devoted to a summary of this work.

\section{A Phenomenological Wind Model}\label{sec:method}

We adopt a phenomenological model to describe the mass accretion rate profile $\dot{M}(R)$ of the disc suffering from a wind
\begin{equation}\label{eq:mass_rate}
  \dot{M}(R)=\dot{M}_{\rm in}\left(\frac{R}{R_{0}}\right)^{s}, \ \ R \geq R_{0},
\end{equation}
where $\dot{M}_{\rm in}$ is the mass accretion rate at $R_{0}$ and $R_{0}$ is chosen as the inner edge of the disc where wind can dominate over inflow. The wind parameter $s$ is kept as constant in the disc, and the no wind special case is at $s=0$.
Numerical simulations and theoretical works for the hot accretion flows suggested that
$R_{0}\simeq 20-40\ R_{\rm g}$ (\citealt{Yuan2012a,Narayan2012,Yuan2015}), where $R_{\rm g}=GM_{\rm BH}/c^2$ is the gravitational radius of a black hole, $G$ is the gravitational constant, $c$ is the speed of light, and $M_{\rm BH}$ is the black hole mass. However, different wind production mechanisms for the thin disc can result in different $R_{0}$.
We first adopt $R_{\rm 0}=6\ R_{\rm g}$, which is the the innermost stable circular orbit for a Schwarzschild black hole.
The value of $R_{0}$ depends on the wind launching mechanism. For the line driven wind, $R_{0}$ is related to the UV emitting photon of the disc, and the launching radius could be on the scale of $\sim100~R_{\rm g}$ (e.g., \citealt{Proga2000,Risaliti2010}). The wind launching radius of the thermally driven wind is even larger. However, if we consider magnetic field, magnetically driven winds can produce high-velocity wind from the very inner region of the accretion disc, with  the launching radius being a few $R_{\rm g}$ (e.g., \citealt{Blandford1982,Contopoulos1994,Konigl1994,Fukumura2018}, and references therein).  For example,  \citet{Fukumura2018} find that the magnetically driven wind can originate around $\sim10~R_{\rm g}$. In the observational side,  X-ray observations of ultrafast outflows via blue-shifted absorption lines suggest the upper limit of the wind launching radius being $\sim20~R_{\rm g}$ \citep{Tombesi2012,Tombesi2015}.
In addition, we have discussed the effect of different $R_{0}$ on the wind-corrected disc size in the appendix. We find that a reasonably larger $R_{0}$ has a weak effect on our results.

With the radius-dependent mass accretion rate, the effective temperature profile $T(R)$ for a thermally radiating black body disc can be obtained by
\begin{equation}\label{eq:flux}
  \frac{3GM_{\rm BH}\dot{M}(R)}{8\pi R^{3}}=\sigma T^{4}(R),
\end{equation}
which leads to $T(R)\propto R^{-\beta}$ with $\beta=(3-s)/4$. , where $\sigma$ is the Stefan-Boltzmann constant. For simplicity, we have ignored the correction factor $1-(R_{0}/R)^{1/2}$ of the inner edge. The effect is minor once the disc size is much larger than $R_{0}$. We will discuss the effect of this simplification below. Such a flatter temperature profile due to the disc wind has been studied in previous works (e.g., \citealt{Witt1997,Knigge1999,Laor2014}, also see \citealt{Sun2018b} for a very recent work.).
Due to the angular momentum transport by the wind, \citet{Laor2014} find that inserting $\dot{M}(R)$ instead of a constant $\dot{M}$ cannot quantify the effect of disc wind on the radiation by using a different radial-dependent mass flux for the accretion rate. With the wind prescription in this work and solving their Equations~(27$-$28) to obtain the local flux $F(R)$, we find that such a temperature profile modification is still a good proximation if we neglect the inner edge effect. However, after the further consideration of the vertical angular momentum transfer by the wind, the modification could be significant. This will be discussed in Section~\ref{sec:ang}.

The resulting surface brightness at a rest wavelength $\lambda_{0}$ is given by
\begin{equation}\label{eq:fv}
  f_{\nu}=\frac{2h_{\rm p}c}{\lambda^3_{0}}\left[\exp\left(\frac{R}{R_{\lambda_{0}}}\right)^{\beta}-1\right]^{-1},
\end{equation}
where the scale length $R_{\lambda_{0}}$ characterized by $kT(R)=h_{\rm p}c/\lambda_{0}$ defines the theory size of the disc,
\begin{equation}\label{eq:rth}
  R_{\lambda_{0},{\rm th}}(\beta)=\left[\frac{45G\lambda_{0}^{4}M_{\rm BH}\dot{M}_{\rm in}}{16\pi^{6}h_{\rm p}c^{2}R_{0}^{s}}\right]^{1/(3-s)}.
\end{equation}
The size scales with the wavelength $\lambda$ as $R_{\lambda,{\rm th}}\propto \lambda^{1/\beta}$. Here $h_{\rm p}$ is the Planck constant, and $k$ is the Boltzmann constant. Assuming a luminosity $L=\eta \dot{M}_{\rm in}c^{2}$ and an Eddington ratio of $L/L_{\rm Edd}=f$, where $\eta$ is the radiative efficiency, and $L_{\rm Edd}$ is the Eddington luminosity, we can rewrite Equation~(\ref{eq:rth}) as
\begin{equation}\label{eq:rth2}
  R_{\lambda_{0},\rm th}(\beta)=\left[\frac{45fG^{2}m_{\rm p}\lambda_{0}^{4}M_{\rm BH}^{2}}{4\eta\pi^{5}h_{\rm p}c^{3}\sigma_{\rm T}R_{0}^{s}}\right]^{1/(3-s)},
\end{equation}
where $\sigma_{\rm T}$ is the Thomson cross-section, $m_{\rm p}$ is the mass of a proton. We adopt a typical Eddington ratio $f=0.1$ for quasars \citep{Shen2008}, although \citet{Kollmeier2006} estimate a slightly larger value of $f\simeq1/4$. The dependence of $R_{\lambda_{0},{\rm th}}$ on the black hole mass $M_{\rm BH}$ is modified as $R_{\lambda_{0},{\rm th}}\propto M_{\rm BH}^{(2-s)/(3-s)}$, which recoveries to the predicated slope of $2/3$ from the thin-disc theory in the case of the no-wind model (i.e., $s=0$).

\begin{figure}
%\vbox to3.2in{\rule{0pt}{3.2in}} \special{psfile=fig1.PDF voffset=0 hoffset=0 vscale=80 hscale=80 angle=0}
\centering
\includegraphics[width=0.45\textwidth]{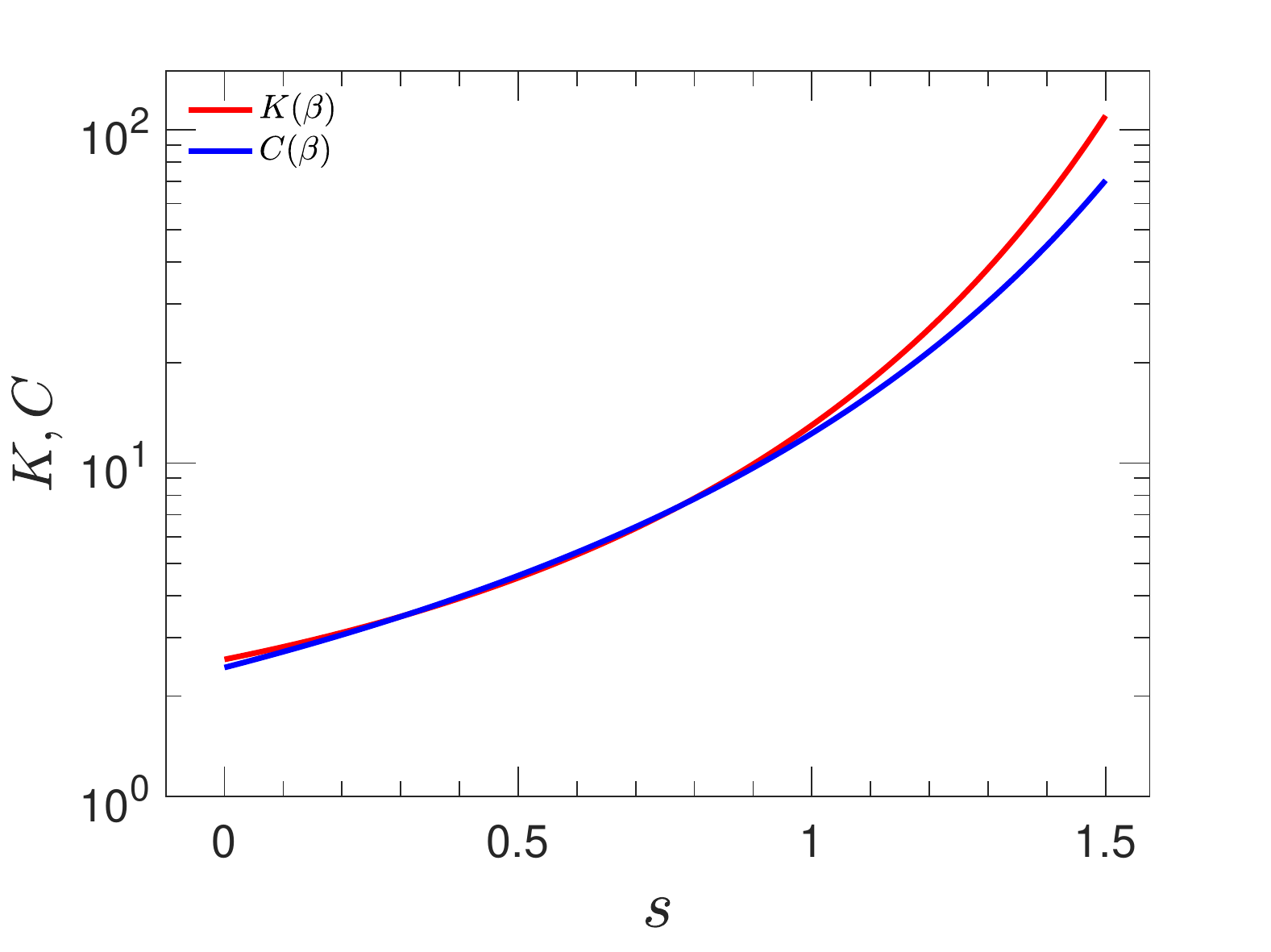}
\caption{$K(\beta)$ (red) and $C(\beta)$ (blue) as functions of the wind parameter $s$.} \label{fig:kc}
\end{figure}

Under the same model assumption, we can obtain another disc size by setting the integrated surface brightness profile over the whole disc with the magnification-corrected quasar fluxes at a given band. Here the relativistic effect is ignored for simplicity because this effect is unimportant for the optical-UV emission. The flux size under the wind model is then given by
\begin{eqnarray}\label{eq:rflux}
  R_{\lambda_{0},\rm flux}(\beta)&=&\frac{D_{\rm OS}}{\sqrt{4\pi h_{\rm p}c\cos i}}\sqrt{\frac{K(3/4)}{K(\beta)}}\lambda_{0}^{3/2}F_{\nu}^{1/2} \nonumber \\
  &=& \frac{2.8\times10^{15}}{h\sqrt{K(\beta)/K(3/4)\cos i}}\frac{D_{\rm OS}}{r_{\rm H}}\left(\frac{\lambda_{0}}{\mu \rm m}\right)^{3/2} \nonumber \\
  &\times&\left(\frac{\rm zpt}{2409\ \rm Jy}\right)^{1/2}10^{-0.2(m-19)}\ {\rm cm},
\end{eqnarray}
where $i$ is the inclination angle of the disc, $m$ is the magnification-corrected magnitude for microlensing sources, $D_{\rm OS}/r_{\rm H}$ is the source angular distance in units of the Hubble radius $r_{\rm H}\equiv c/H_{0}$, $h=H_{0}/(100\ {\rm km\ s^{-1}\ Mpc^{-1}})$, and for $kT(R_{\rm out})/h_{\rm p}\ll c/\lambda_{0}\ll kT(R_{0})/h_{\rm p}$,
\begin{equation}\label{eq:kbeta}
  K(\beta)=\int_{0}^{\infty}\left[\exp(x^{\beta})-1\right]^{-1}x{\rm d}x.
\end{equation}
It clearly shows that the flux size of the disc $R_{\lambda_{0},\rm flux}$ depends on the temperature profile via the wind parameter $s=3-4\beta$. For the no-wind model, $K(\beta=3/4)=2.58$. The dependence of $K$ on the wind strength $s$ is shown in Figure~\ref{fig:kc}, which implies that $R_{\lambda_{0},{\rm flux}}$ decreases significantly with the wind strength.

For the  disc-size measurement via microlensing variability, it is the half-light radius of the gravitationally lensed quasars that is measured. Based on the wind-corrected surface brightness profile in Equation~(\ref{eq:fv}), we can relate the half-light radius of the disc $R_{\lambda_{0},{\rm half}}$ with the scale size $R_{\lambda_{0}}$ in Equation~(\ref{eq:fv}) as
\begin{equation}\label{eq:rmic}
  R_{\lambda_{0},{\rm half}}(\beta)=C(\beta)R_{\lambda_{0},{\rm mic}}(\beta),
\end{equation}
where $C(\beta)$ is the conversion factor from $R_{\lambda_{0}}$ in Equation~(\ref{eq:fv}) to $R_{\lambda_{0},{\rm half}}$. Here $R_{\lambda_{0}}$ is labeled as the third disc size, namely the microlensing size $R_{\lambda_{0},{\rm mic}}$, which can be directly compared with other disc size measurements. $C(\beta)$ is defined as
\begin{equation}\label{eq:cbeta}
  C(\beta)=\mathcal{F}^{-1}\left[\frac{1}{2}K(\beta)\right],
\end{equation}
where $\mathcal{F}^{-1}(\cdot)$ is the inverse function of
$$\mathcal{F}(x)=\int_{0}^{x}\left[\exp(u^{\beta}) -1\right]^{-1}u{\rm d}u.$$ Under the no-wind circumstance, i.e., $\beta=3/4$, $C(3/4)=2.44$. We also plot the profile of $C$ as a function of the wind parameter $s$ in Figure~\ref{fig:kc}. The half-light radius measured from microlensing is nearly model independent \citep{Mortonson2005,Congdon2007}, so a different wind strength can modify the disc size $R_{\lambda_{0},{\rm mic}}$ as well.

Theoretically speaking, the three disc sizes $R_{\lambda_{0},{\rm th}}$, $R_{\lambda_{0},{\rm flux}}$ and $R_{\lambda_{0},{\rm mic}}$ defined in Equations~(\ref{eq:rth2},\ref{eq:rflux},\ref{eq:rmic}), respectively, should be
consistent with each other. However, there are striking discrepancies among these three size measurements observationally. Even though the offset between the microlensing size measurements (Equation~\ref{eq:rmic}) and theory size (Equation~\ref{eq:rth2}) tends to be somewhat smaller, which can be attributed by other uncertainties (e.g., radiative efficiency $\eta$, and Eddington ratio $f$), the discrepancy between the microlensing size (Equation~\ref{eq:rmic}) and the expectation from the flux measurement (Equation~\ref{eq:rflux}) is more significant in most microlensed quasars with the former being larger than the latter by $0.6\pm0.3$ dex \citep[e.g.,][]{Pooley2007,Poindexter2008,Morgan2010,Morgan2012}.

\begin{figure}
%\vbox to3.2in{\rule{0pt}{3.2in}} \special{psfile=fig1.PDF voffset=0 hoffset=0 vscale=80 hscale=80 angle=0}
\centering
\includegraphics[width=0.45\textwidth]{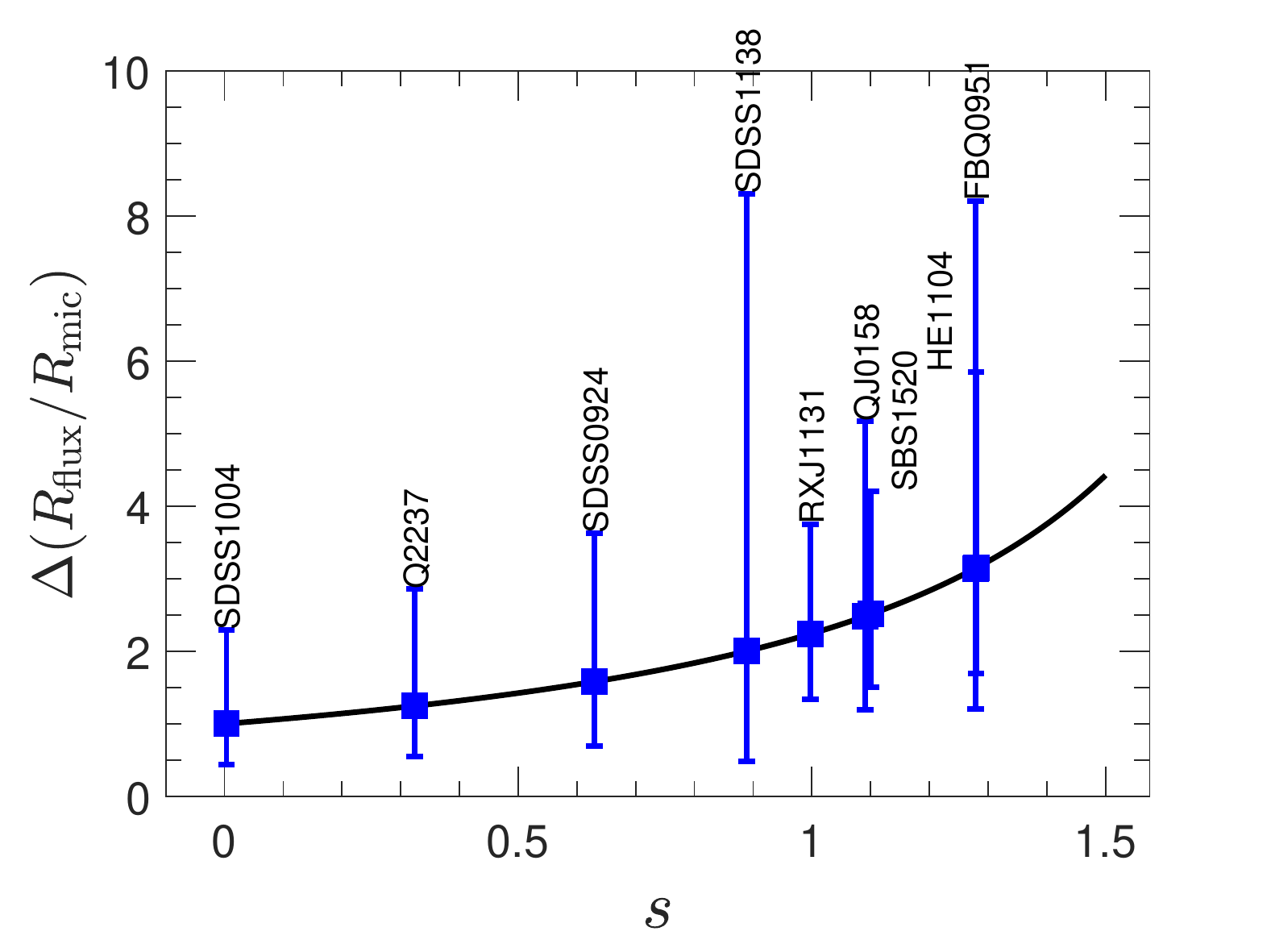}
\caption{The correction factor $\Delta(R_{\rm flux}/R_{\rm mic})$ as a function of wind parameter $s$. The observational disc-size ratio discrepancies $R_{\rm mic,obs}$/$R_{\rm flux,obs}$ for several quasars with their error bars are obtained via the Monte Carlo sampling. The inferred wind strength $s$ determined by setting $\Delta=R_{\rm mic,obs}$/$R_{\rm flux,obs}$ are shown with blue squares.} \label{fig:size_ratio}
\end{figure}

We mainly focus on the major size problem arisen from the discrepancy between the flux size $R_{\lambda_{0},{\rm flux}}$ and the microlensing one $R_{\lambda_{0},{\rm mic}}$. As both $R_{\lambda_{0},{\rm flux}}$ and $R_{\lambda_{0},{\rm mic}}$ are sensitive to the wind parameter $s$, the correction factor of the size ratio due to the disc wind based on Equations~(\ref{eq:rflux}) and (\ref{eq:rmic}) can be described by
\begin{eqnarray}\label{eq:size_ratio}
  \Delta(R_{\rm flux}/R_{\rm mic})&\equiv& \frac{R_{\lambda_{0},{\rm flux}}(\beta)}{R_{\lambda_{0},{\rm flux}}(3/4)}\frac{R_{\lambda_{0},{\rm mic}}(3/4)}{R_{\lambda_{0},{\rm mic}}(\beta)} \nonumber \\
  &=&\sqrt{\frac{K(3/4)}{K(\beta)}}\times\frac{C(\beta)}{C(3/4)},
\end{eqnarray}
which is shown as the solid line in Figure~\ref{fig:size_ratio}. It clearly demonstrates that the flux to microlensing size ratio correction can be a factor of $\sim3$ as the wind parameter increases to $\sim1.3$. It, therefore, indicates that a disc model with a strong wind can potentially resolve the underestimation of flux size measurements up to 0.6 dex.

\section{Influence of Angular Momentum Transfer by Wind}\label{sec:ang}

While our phenomenological treatment of the wind described by Equation~(\ref{eq:mass_rate}) is the standard treatment in most previous works, wind could play another role in modifying the inflow by transferring the angular momentum outward. This has been studied for two types of disc models. \citet{Xie2008} have investigated this problem in details for the advection-dominated accretion flow by incorporating the interchange of mass, momentum, and energy between the inflow and outflow. They confirm that the phenomenological treatment of wind is reasonable for the hot accretion flow.
For the thin disc, \citet{Kuncic2007} proposed a phenomenological model by including the vertical angular momentum transfer induced by the wind (see \citealt{Laor2014} for a different phenomenological model for the thin disc wind). This provides a straightforward way to be compared with our wind model.

For the purpose of reconciling the disc size discrepancy, it is related with the modification of the radial temperature profile of the disc determined by the radiative flux, which is required to balance with the viscous dissipation rate.
The radiative flux of a thin disc after the consideration of the angular momentum transferred by the wind can be expressed as \citep{Kuncic2007}
\begin{equation}\label{eq:flux_ang}
  F_{+}(R)=\frac{3GM_{\rm BH}\dot{M}(R)}{8\pi R^{3}}\xi_{\rm corr}(R).
\end{equation}
The additional radial-dependent correction factor $\xi_{\rm corr}(R)$ is given by
\begin{eqnarray}\label{eq:xi}
  \xi_{\rm corr}(R)&=&1-\left(\frac{R}{R_{0}}\right)^{-\frac{1}{2}-s}-\frac{1+2s}{1+2s-2w} \nonumber \\
  &\times& \left(\frac{R}{R_{0}}\right)^{-w}\left[1-\left(\frac{R}{R_{0}}\right)^{-\frac{1}{2}-s+w}\right],
\end{eqnarray}
where $w$ is another parameter describing the radial dependence of the angular momentum transfer by the wind \citep{Kuncic2007}.  A smaller $w$ indicates a stronger effect of angular momentum transportation by wind.  For the case of $s=0$ and $w\rightarrow\infty$,  $\xi_{\rm corr}(R)$ recoveries to $1-(R/R_{0})^{-1/2}$, which is the correction factor of the disc inner edge for the model without disc wind\footnote{This correction factor also shows a positive gradient with radius, which can then lead to a flatter temperature profile as we discuss below.}.  There exists no observational and theoretical constraints for this parameter, which makes a detailed quantitative assessment of this effect difficult. However, we can qualitatively discuss this effect in the disc temperature profile.

\begin{figure}
%\vbox to3.2in{\rule{0pt}{3.2in}} \special{psfile=fig1.PDF voffset=0 hoffset=0 vscale=80 hscale=80 angle=0}
\centering
\includegraphics[width=0.45\textwidth]{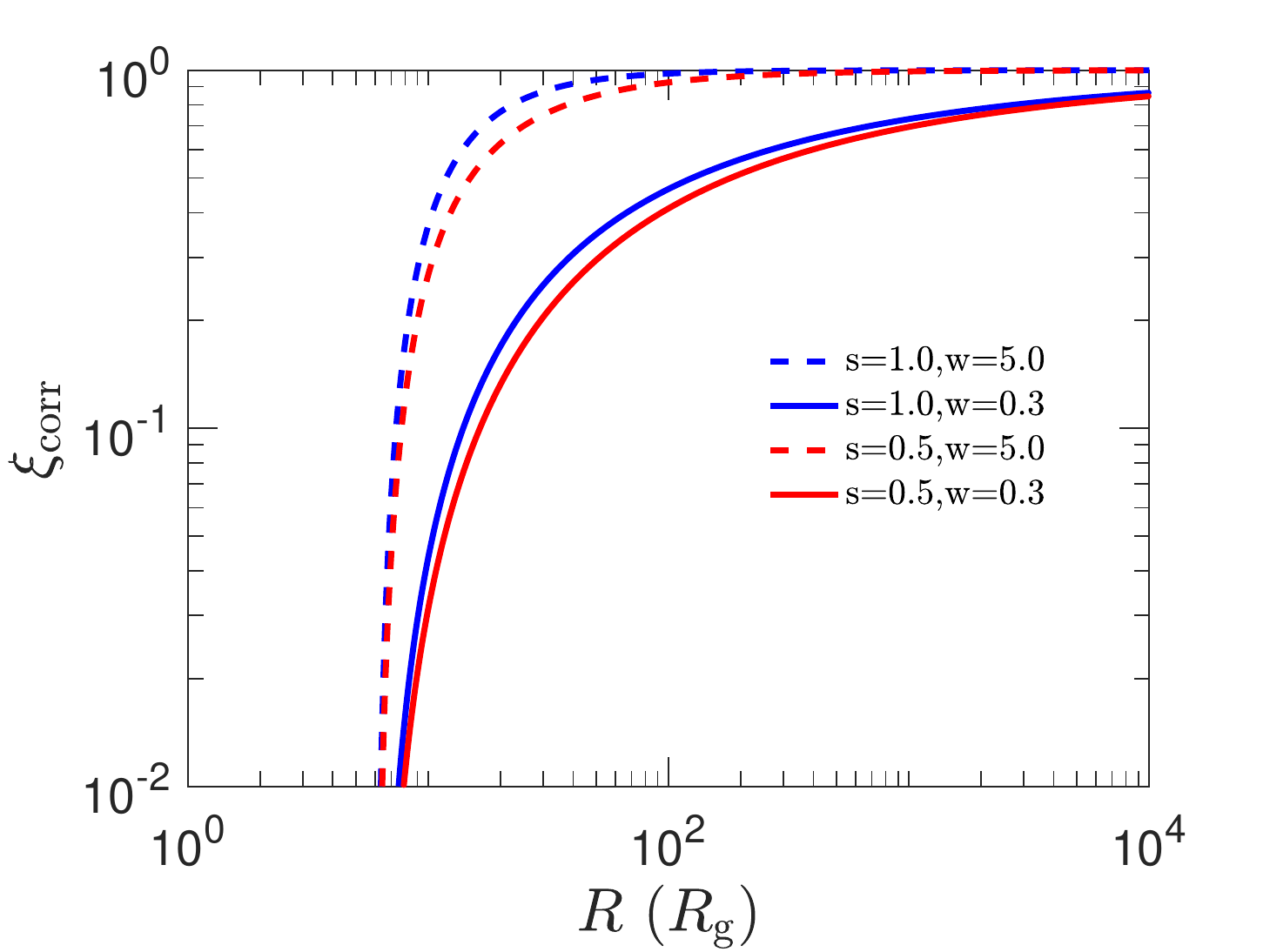}
\includegraphics[width=0.45\textwidth]{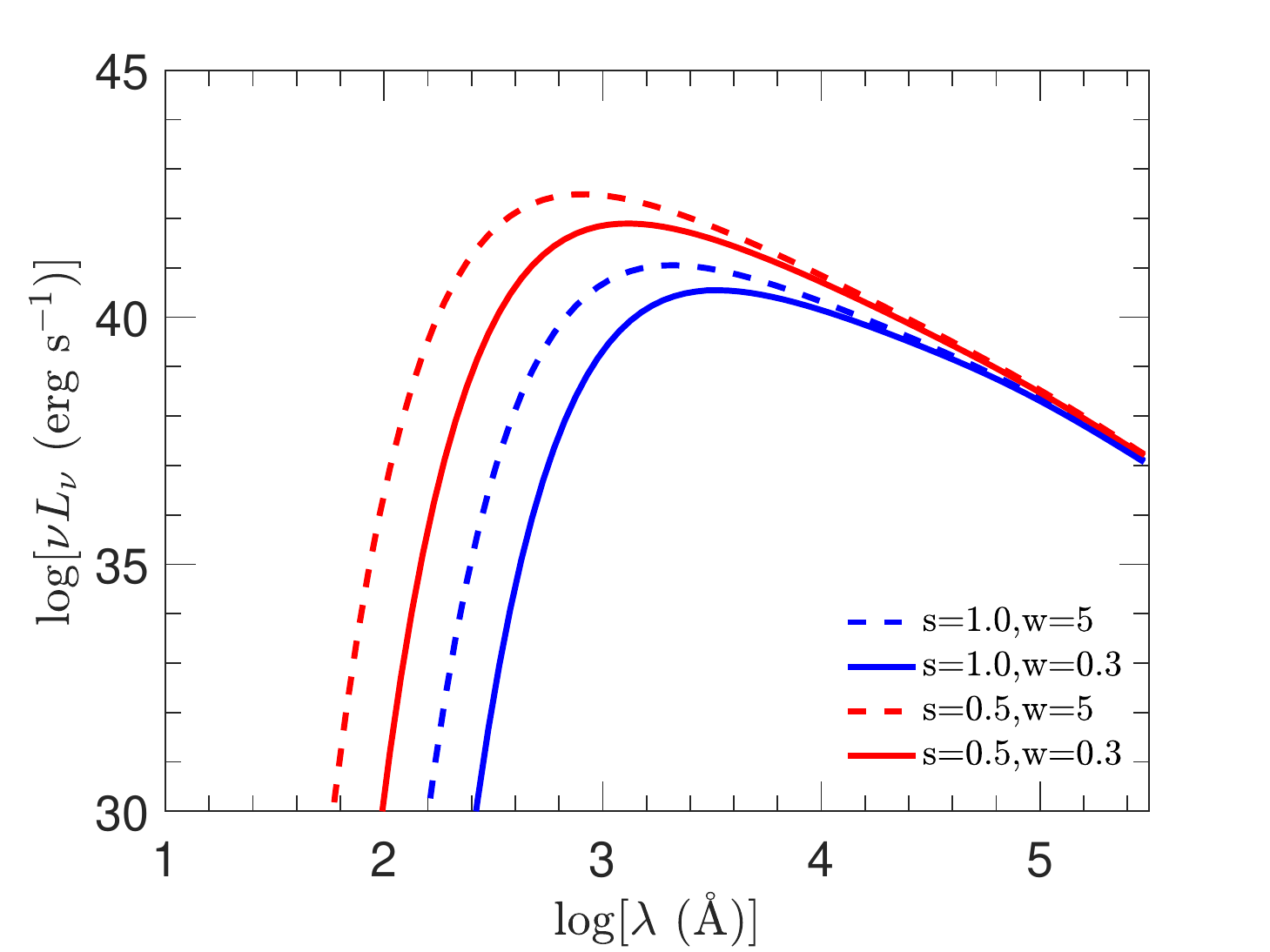}
\caption{Upper panel: the correction factor $\xi_{\rm corr}$ due to the angular momentum transfer by the disc wind for different wind strength $s$ and angular momentum parameter $w$. Lower panel: the corresponding SED of the disc. We adopt a typical black hole mass of $M_{\rm BH}=10^{8}~M_{\odot}$, an Eddington ratio of $f=0.1$, and a radiative efficiency of $\eta=0.1$.} \label{fig:xi}
\end{figure}

We show the radial profile of $\xi_{\rm corr}(R)$ for different combinations of $s$ and $w$ parameters in the upper panel of Figure~\ref{fig:xi}. Since a steep positive radial gradient of $\xi_{\rm corr}(R)$ can be obtained in a large radial range of the disc, and the effective temperature is determined by $T(R)\propto F_{+}^{1/4}(R)$, which in turn results in an even flatter temperature profile. Furthermore, the slope becomes steeper in the inner regions of the disc.
This is equivalent to a larger effective wind parameter $s_{\rm eff}$ in the inner region. Therefore, it can help to resolve the disc size problem with a relatively small wind parameter.

Before applying our wind model to several microlensed quasars, we discuss how the disc wind can modify the emitting spectrum.
Another important problem of the standard thin disc model is that the theory cannot explain quite well a universal characteristic turnover around $1000~{\AA}$ of SEDs in AGNs, indicating a maximal disc temperature of $\sim50000$ K. A wind scenario has been proposed and studied in details by \citet{Laor2014} to resolve this problem (see also \citet{Kuncic2007,Slone2012} for similar works). A very similar wind model has also recently been applied to NGC 5548 to resolve the inter-band time lag and SED problems \citep{Sun2018b}.
As our wind model share some similarities with previous works, we expect our model can also be used to solve this SED problem as well. To illustrate this point, we have added one plot in the lower panel of Figure~\ref{fig:xi} showing the modified SED from the disc with wind. It can be seen that a disc with the wind parameter $s\gtrsim0.5$ after taking into account the angular momentum transport by wind can easily reproduce the observed peak around $\sim1000~\AA$.
This is simply because the temperature profile $T(R)\propto R^{-\beta}$ becomes shallower due to the existence of disc wind. As a result, the maximal temperature  reached in the inner region is decreased to a few $\times10^4$ K.

\section{Application to Microlensed Quasars}\label{sec:results}

We first apply our wind model to several gravitationally microlensed quasars without considering the angular momentum transfer effect. \citet{Morgan2010} have collected 11 quasars with their microlensing size and flux size reported. We select 9 sources from their samples as shown in Table~\ref{tab:para}, except that HE 0435$-$1223 and PG 1115$+$080 are excluded. The reason for the exclusion of these two source will be discussed below.

%\begin{verbatim}
\begin{table*}
  %\centering
  \begin{center}
  \caption{Sources parameters and corrected disc sizes} \label{tab:para}
  \resizebox{0.95\textwidth}{!}{
  \begin{tabular}{lccccccc}
     \hline\hline
     % after \\: \hline or \cline{col1-col2} \cline{col3-col4} ...
     Objects & $M_{\rm BH}\ (10^{9}\ M_{\odot})$ & $\lambda_{0}\ (\mu{\rm m})$ & $\log(R_{\rm mic,obs}/{\rm cm})$ & $\log(R_{\rm flux,obs}/{\rm cm})$ & $s$ & $\log(R_{\rm flux,mic}^{\rm corr}/{\rm cm})$ & $\log(R_{\rm th}^{\rm corr}/{\rm cm})$ \\
     \hline
     QJ 0158$-$4325 & 0.16 & 0.310 & $15.6\pm0.3$  & $15.2\pm0.1$ & 1.1 & $14.7\pm0.3$ & 14.9 \\
     SDSS 0924$+$0219 & 0.11 & 0.277 &  $15.0_{-0.4}^{+0.3}$  & $14.8\pm0.1$ & 0.6 & $14.6\pm0.4$ & 14.7 \\
     FBQ 0951$+$2635 & 0.89 & 0.313 &  $16.1\pm0.4$ &  $15.6\pm0.1$ & 1.3  & $14.9\pm0.4$ & 15.3 \\
     SDSS 1004$+$4112 & 0.39 & 0.228  & $14.9\pm0.3$  & $14.9\pm0.2$ &  0.02 & $14.9\pm0.3$ & 14.9 \\
     HE 1104$-$1805 & 2.37 & 0.211 &  $15.9_{-0.3}^{+0.2}$  & $15.4\pm0.1$ & 1.3  & $14.9\pm0.3$ & 15.5 \\
     RXJ 1131$-$1231 & 0.06 & 0.400  & $15.2\pm0.2$  & $14.8\pm0.1$ &  1.0 & $14.2\pm0.2$& 14.7 \\
     SDSS 1138$+$0314 & 0.04 & 0.203  & $14.9\pm0.6$  & $14.6\pm0.1$ &  0.9 & $14.4\pm0.6$ & 14.5 \\
     SBS 1520$+$530 & 0.88 & 0.245 &  $15.7\pm0.2$  & $15.3\pm0.1$ & 1.1  & $14.9\pm0.2$ & 15.3 \\
     Q2237$+$030 & 0.90 & 0.208 &  $15.6\pm0.3$  & $15.5\pm0.2$ & 0.3  & $15.5\pm0.3$ & 15.2 \\
     \hline\hline
  \end{tabular}
  }
  \end{center}
 \begin{minipage}{16cm}
 {NOTE: Data are collected from \citet{Morgan2010}, except for QJ 0158$-$4325 \citep{Morgan2012} and RXJ 1131$-$1231 \citep{Dai2010}. $R_{\rm flux,obs}$ and $R_{\rm mic,obs}$ are disc sizes in the no-wind model. The typical uncertainty of inferred $s$ is $\sim0.5$, as seen from Figure~\ref{fig:size_ratio}. $R_{\rm flux,mic}^{\rm corr}$ is the corrected flux/microlensing size after the correction based on our wind model measured at $0.25\ \mu{\rm m}$. The  $1\sigma$ errors of $R_{\rm flux,mic}^{\rm corr}$ are obtain by 5000 Monte Carlo sampling. Note that the corresponding half-light radius corrected by $C(\beta)$ are larger by a factor of ten on average.   $R_{\rm th}^{\rm corr}$ is the wind-modified theory disc size with $\eta=0.1$ and $f=0.1$ at $0.25\ \mu{\rm m}$ (Equation~\ref{eq:rth2}). The typical uncertainty of $R_{\rm th}^{\rm corr}$ is $\sim0.1$ dex.   \\
 }
 \end{minipage}
\end{table*}
%\end{verbatim}

To reconcile between the flux and microlensing size measurements, one can make the observational size ratio $R_{\rm mic,obs}/R_{\rm flux,obs}$ be linked to the correction factor defined in Equation~(\ref{eq:size_ratio}). The observational size ratios and the $1\sigma$ uncertainties for all sources can be obtained through a Monte Carlo sampling of $R_{\rm flux,obs}$ and $R_{\rm mic,obs}$ as listed in Table~\ref{tab:para}. By setting $R_{\rm mic,obs}/R_{\rm flux,obs}=\Delta(R_{\rm flux}/R_{\rm mic})$, we can obtain the required wind parameters, which are shown in Figure~\ref{fig:size_ratio} and listed in Table~\ref{tab:para}. It indicates that the underestimation of the flux size can be well explained with a wind strength $s\sim1$ for our sources. The typical uncertainty of $s$ is relatively large ($\sim0.5$), which is due to the large uncertainty of $R_{\rm mic,obs}/R_{\rm flux,obs}$ shown in Figure~\ref{fig:size_ratio}. In addition, the required $s$ values can be decreased by $\sim0.1$ after considering the inner edge effect in Equations~(\ref{eq:kbeta}) and (\ref{eq:cbeta}). For instance, such correction can be considered by estimating, e.g., $R_{0}\sim0.5R_{\lambda_{0}}$. Then the lower limit of the integral for $K(\beta)$ can be set to be 0.5, and another correction term $(1-\sqrt{0.5/x})^{-1/4}$ is included in the integral of $K(\beta)$ and $C(\beta)$, e.g., $K(\beta)=\int_{0.5}^{\infty}\left[\exp(x^{\beta}(1-\sqrt{0.5/x})^{-1/4})-1\right]^{-1}x{\rm d}x$. The inner edge effect for the theory size in Equation~(\ref{eq:rth2}) is $\lesssim20\%$ by inserting the factor of $1-(R/R_{0})^{-1/2}$ into Equation~(\ref{eq:flux}).
It is thus very likely to interpret the size discrepancy with a wind strength $s<1.0$ after considering those uncertainties.
We note that \citet{Laor2014} argue that their disc wind model cannot solve the disc size problem. This is likely because our wind mass flux is stronger than theirs, which can then result in a flatter temperature profile, based on the different radial dependence of the wind.

Observationally, the wind parameter $s$ for the thin disc has not been well constrained up to date. There are some constraints for the hot accretion flow in several low-luminosity AGNs [e.g., $s\sim0.5$ for M87 (\citealt{Russell2015}) and NGC 3115 (\citealt{Wong2014})] and Sagittarius A$^\star$ ($s\sim1.0$; \citealt{Wang2013}) by X-ray observations. Numerical simulations of hot accretion flows show that $s\sim0.5-1.0$ (see review by \citealt{Yuan2014} and references in the Introduction.).

More importantly, the required wind parameter to interpret the size discrepancy can be reduced by $\sim0.5$ if the angular momentum transfer effect by the wind is considered, as estimated from the upper panel of Figure~\ref{fig:xi}. This can make the required $s<1.0$ (e.g., FBQ 0951$+$2635, HE 1104$-$1805) and relieve the disc size problem for some sources, e.g., HE 0435$-$1223 and PG 1115$+$080.
For  HE 0435$-$1223 and PG 1115$+$080, their size discrepancies are 0.8 and 1.5 dex, respectively. We find that a very large wind parameter ($s>1.5$) has to be adopted to account for the disc size discrepancy if we only consider the inward decrease of the mass accretion rate. An additional consideration of significant vertical angular momentum transport by wind  (i.e., small $w$ in Equation~\ref{eq:xi}), which is physically very likely, can solve this problem.  Another possible solution is to consider the radiation contamination from larger scales such as the broad-line region and/or scattering of the disc flux on larger physical scales \citep{Morgan2010}, since these external effects can potentially reduce the intrinsic size discrepancy.
Considering these complexities, we think that the wind parameters $s$ as shown in Figure~\ref{fig:size_ratio} are in the reasonable range.

\begin{figure}
%\vbox to3.2in{\rule{0pt}{3.2in}} \special{psfile=fig1.PDF voffset=0 hoffset=0 vscale=80 hscale=80 angle=0}
\centering
\includegraphics[width=0.45\textwidth]{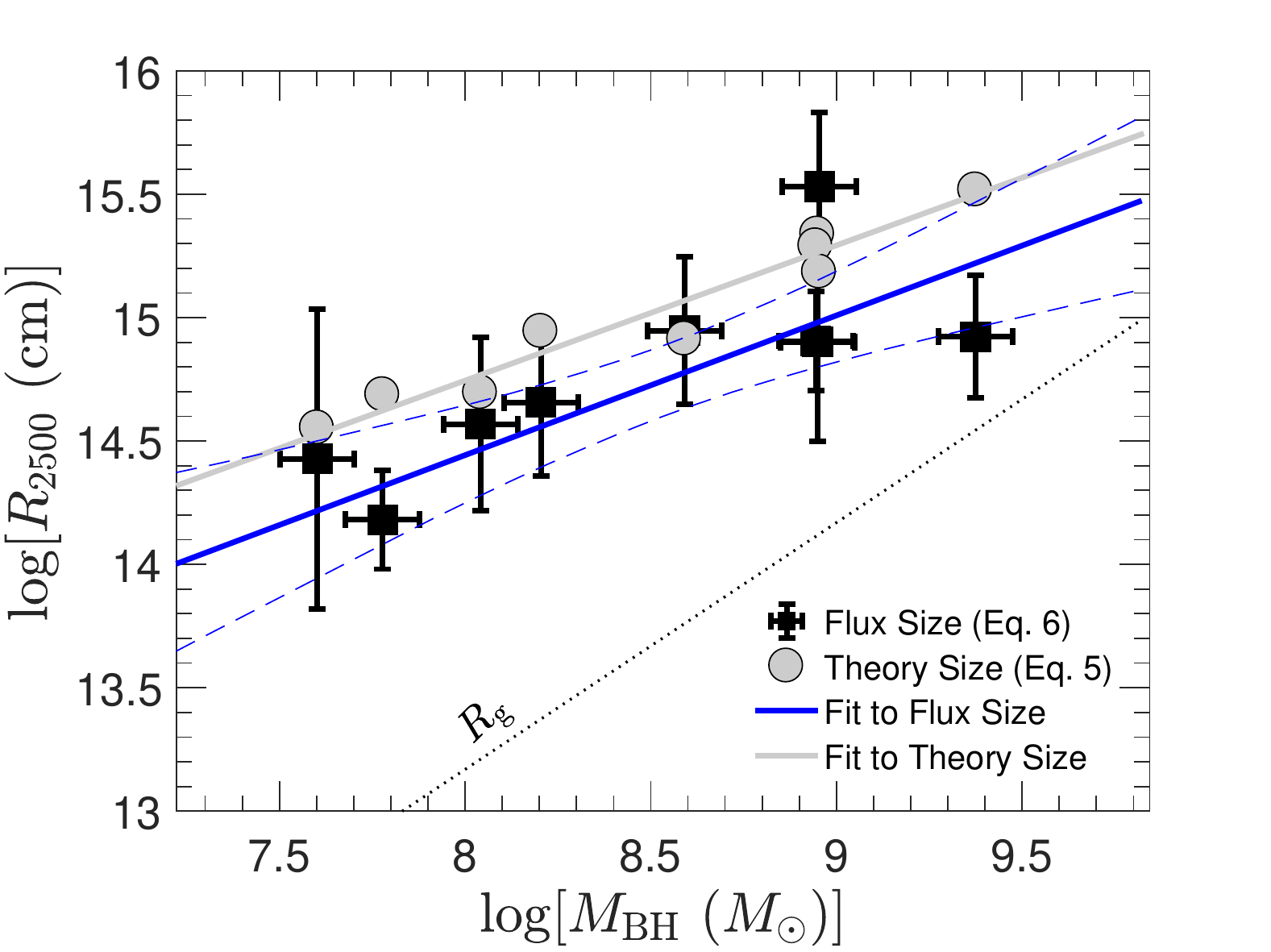}
\caption{Wind-corrected disc size at $\lambda_{0}=0.25\ \mu{\rm m}$ as a function of black hole mass $M_{\rm BH}$. The black squares with error bars are flux sizes, which have been corrected to match the microlensing sizes. The solid blue line shows the fit to the corrected flux size VS. black hole mass with the dashed blue lines representing the $1\sigma$ uncertainty. The grey circles show the wind-corrected theory sizes from Equation~(\ref{eq:rth2}) with a radiative efficiency of 0.1. The grey line represents the fit to the theory size. The dotted line shows the gravitational radius $R_{\rm g}$ of the black hole.} \label{fig:size_mbh}
\end{figure}

With the inferred wind parameters, we can calculate the wind-corrected disc sizes. Note that the flux and microlensing sizes are already consistent with each other by definition. We show the wind-corrected flux (or microlensing) size $R_{\rm mic}^{\rm corr}$ at $0.25\ \mu{\rm m}$ as a function of the black mass $M_{\rm BH}$ as black squares in Figure~\ref{fig:size_mbh}. The $1\sigma$ uncertainties of the modified microlensing sizes are obtained via Monte Carlo simulations as well. The typical black hole mass uncertainty is 0.1 dex.
A Markov Chain Monte Carlo method \citep{Lewis2002,Li2015} is applied to fit the correlation between $R_{\rm mic}^{\rm corr}$ and $M_{\rm BH}$ by including the uncertainties for both variables yielding

\begin{equation}\label{eq:size_mbh}
  \log\left(\frac{R_{\rm mic}^{\rm corr}}{\rm cm}\right)=(9.91\pm1.45)+(0.56\pm0.17)\log\left(\frac{M_{\rm BH}}{M_{\odot}}\right).
\end{equation}
The corresponding best fit and $1\sigma$ error band are shown as solid and dashed blue lines, respectively, with the best-fitted statistics $\chi^{2}_{\nu}=6.0/7\simeq0.9$, suggesting a reasonable fit to the data.

By assuming a typical radiative efficiency of $\eta=0.1$ and an Eddington ratio $f=0.1$ for all sources, the wind-corrected thin-disc theory sizes $R_{\rm th}^{\rm corr}$ as defined in Equation~(\ref{eq:rth2}) can be obtained, which are represented as grey circles in Figure~\ref{fig:size_mbh}.
Assuming an uncertainty of $0.1$ dex for both $R_{\rm th}^{\rm corr}$ and $M_{\rm BH}$, a power-law function is applied to fit between $R_{\rm th}^{\rm corr}$ and $M_{\rm BH}$, which leads to
\begin{equation}\label{eq:rth_mbh}
  \log\left(\frac{R_{\rm th}^{\rm corr}}{\rm cm}\right)=(10.4\pm0.6)+(0.55\pm0.07)\log\left(\frac{M_{\rm BH}}{M_{\odot}}\right).
\end{equation}
The fitted power law is shown as the grey line in Figure~\ref{fig:size_mbh} with $\chi^{2}_{\nu}=4.6/7\simeq0.7$, suggesting a slightly overestimate of uncertainties for $R_{\rm th}^{\rm corr}$. The power-law function in Equations~(\ref{eq:size_mbh}) and (\ref{eq:rth_mbh}) are roughly consistent with each other after considering the $1\sigma$ uncertainties, suggesting that the three disc sizes can now be in agreement with each other within the framework of our wind model. We can also judge the consistence between $R_{\rm th}^{\rm corr}$ and $R_{\rm mic}^{\rm corr}$ by calculating $\chi_{\nu}^{2}$ between these two disc sizes, which gives $\chi_{\nu}^{2}\simeq1.7$. The slightly large $\chi_{\nu}^{2}$ is mainly contributed by two sources, RXJ 1131$-$1231 and HE 1104$-$1805. If these two sources are excluded from the test above, we obtain $\chi_{\nu}^2\simeq0.9$, confirming a general consistency between these disc sizes after corrected by the wind.

After considering the angular momentum transfer effect as discussed in Section~\ref{sec:ang}, which reduces the wind parameter $s$ required, these two sizes will further shift toward each other (e.g., RXJ 1131$-$1231). In addition, a slightly higher radiative efficiency (e.g., $\eta\simeq0.15$) can also make $R_{\rm th}^{\rm corr}$ closer to $R_{\rm flux}^{\rm corr}$. The power-law index in Equation~(\ref{eq:rth_mbh}) favors a slightly flatter correlation slope, consistent with a wind parameter $s>0$ as suggested by Equation~(\ref{eq:rth2}).

Interestingly, as the above consistency is based on an assumption of radiative efficiency $\eta=0.1$, it implies the reasonability of a canonical value $\eta=0.1$ expected from the thin disc theory \citep{Frank2002}, higher than those estimated from \citet{Morgan2010}. This is because the radiative efficiency should become higher to produce the same flux with a smaller corrected disc size.

As shown in Figure~\ref{fig:size_mbh}, the corrected disc sizes are $\sim10-50\ R_{\rm g}$. These sizes are increased by a factor of $\sim10$ on average when converting them into the half-light radius, which corresponds to the region where most of emission comes from. This suggests that the emission extends to a more diffuse region. It simply lies in the fact that the radial temperature profile $T(R)\propto R^{-\beta}$ becomes flatter due to the existence of wind. A larger disc size also validates our simplification of neglecting inner edge factor $1-(R_{0}/R)^{1/2}$.

Some observations have revealed that the temperature profiles are even steeper than $3/4$ \citep[e.g.,][]{Eigenbrod2008,Jimenez2014,Munoz2016,Motta2017} or just consistent with 3/4 of the standard thin disc \citep[e.g.,][]{Edelson2015,Fausnaugh2018}, which are inconsistent with the predication of our disc wind model. But it should be noted that the observed uncertainties of these inferred temperature slope are still quite large. Considering the large uncertainties of these inferred temperature slope, it is possible that these observations are compatible with our model expectation. More importantly, recent analysis by \citet{Bate2018} found that there are important selection effects that need to be taken into account when using single-epoch microlensing technique to measure accretion disc temperature profiles. They argue that some previous works based on \citet{Jimenez2014}, as well as those in \citet{Motta2017}, likely overestimate the temperature profile slope.  Therefore, we can conservatively conclude that our model is consistent with current observational data.

In the calculation of the radiation from the disc, we do not take account into the radiative transfer in the disc wind. By adopting an average wind parameter $s=0.8$, the wind velocity as the local escape velocity and a viewing angle $45^{\circ}$ for the disc, we estimate the column density of wind as $\sim10^{24}~{\rm cm^{-2}}$. The optical depth through the wind due to the electron scattering opacity is then less than unity. When the wind strength become weaker (e.g., considering the angular momentum transfer effect) and/or the viewing angle becomes smaller, the optical depth contributed by the wind decreases significantly, which suggests that the effect of the wind on the disc radiation can be negligible. In addition, numerical simulations of the thin disc wind from luminous AGNs have found that the column density of the wind is $10^{22}-10^{24}~{\rm cm^{-2}}$  for a wide range of parameter space \citep{Proga2000,Nomura2016,Kraemer2018}, consistent with the values inferred from blueshifted absorption lines \citep{Tombesi2011}. These results suggest that the disc wind is optically thin to the optical-UV radiation both theoretically and observationally and can thus justify our simplified treatment of wind radiative transfer.

\section{Conclusions and Discussions}\label{sec:conclusion}

In this work, we propose a simple ``wind" scenario to resolve the ``size problem" for several microlensed quasars. With a wind strength $s\lesssim1.3$ (where $s$ is defined via $\dot{M}(R)\propto({R}/{R_{0}})^{s}$ ), the temperature profile of the disc becomes  shallower. Our model can thus make three disc sizes, i.e., microlensing size, flux size, and theory size be consistent with each other. In addition to the mass flux carried away by the disc wind, the vertical angular momentum transport by the wind can further help to relieve the observational size mismatch with a smaller wind parameter $s$.

In the meanwhile, the correlation between wind-corrected disc size and black hole mass becomes slightly flatter, which is in agreement with the theoretical expectation from a thin disc suffering from strong wind. With the updated disc size, we find that the radiative efficiency is close to the canonical value of $0.1$ due to a smaller corrected flux size.

Due to the universality of wind in different accretion systems, the microlensing disc size measurements can thus provide a new probe for the wind properties in the inner region of quasars.

\appendix

\section{The effect of different $R_{0}$}

\begin{figure}
%\vbox to3.2in{\rule{0pt}{3.2in}} \special{psfile=fig1.PDF voffset=0 hoffset=0 vscale=80 hscale=80 angle=0}
\centering
\includegraphics[width=0.45\textwidth]{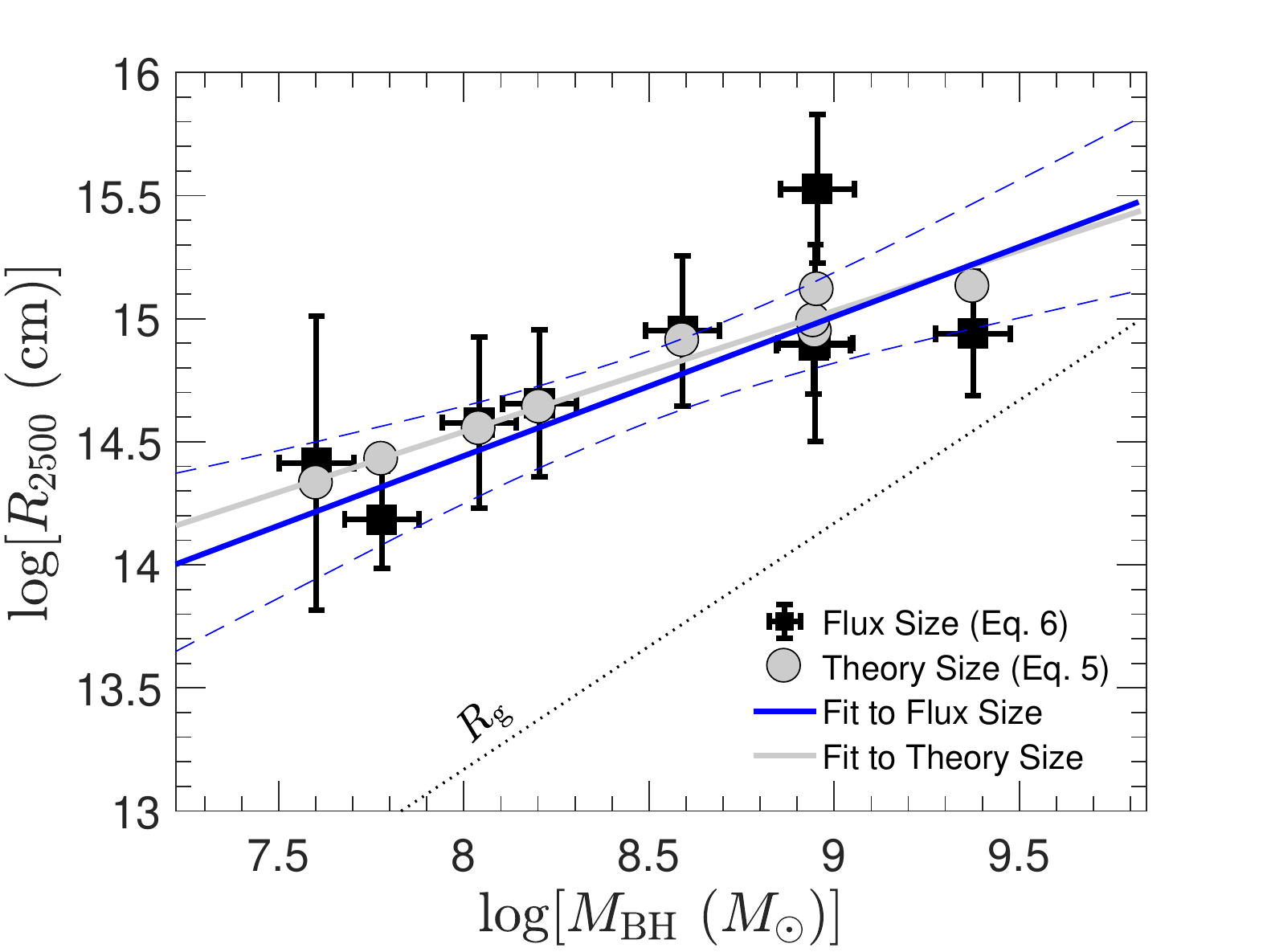}
\caption{Same as Figure~\ref{fig:size_mbh}, but with $R_{0}=20~R_{\rm g}$.} \label{fig:size_mbh_r0}
\end{figure}

Different mechanisms for wind production could result in different launching radii. As an uncertainty for the wind model, we explore the effect of different $R_{0}$ on the disc sizes. A slightly larger $R_{0}=20~R_{\rm g}$ for Equation~(\ref{eq:mass_rate}) is adopted. The wind-corrected disc sizes are shown in Figure~\ref{fig:size_mbh_r0}. With the larger $R_{0}$, the wind-corrected flux size (and microlensing size) does not change significantly as long as the inner edge factor is insignificant, as we have tested above. However, compared with the smaller $R_{0}$ case, the modified theory size (Equation~\ref{eq:rth2}) becomes slightly smaller and more close to the modified flux (microlensing) size. A $\chi^2$ comparison between the two disc sizes (flux size and microlensing size) find that $\chi^2_{\nu}=0.5$.  Therefore, our wind model with a reasonable larger $R_{0}$ can also well resolve the disc size problem.

\section*{Acknowledgments}

We thank the referee for a very constructive report, which has significantly improved the paper. We are grateful to Yaling Jin, Defu Bu, Weixiao Wang and Tim Waters for valuable contributions and comments to this work. This work is supported in part by the National Key Research and Development Program of China (Grant No. 2016YFA0400704), the Natural Science Foundation of China (grants 11573051, 11633006, 11703064, 11650110427, 11661161012), the Key Research Program of Frontier Sciences of CAS (No. QYZDJSSW- SYS008), and Shanghai Sailing Program (grant No. 17YF1422600). This work made use of the High Performance Computing Resource in the Core Facility for Advanced Research Computing at Shanghai Astronomical Observatory and LANL Institutional Computing.

%\software{CosmoMC \citep{Lewis2002} }

\bsp	% typesetting comment
\label{lastpage}

\end{document}